\documentclass[aps,prb,preprint,floatfix,superscriptaddress]{revtex4-1}

\usepackage{graphicx}
\usepackage{dcolumn}
\usepackage{multirow}
\usepackage{amsmath,bm}
\usepackage{xcolor}
\usepackage{pdfcomment}
\usepackage[normalem]{ulem}

\begin{document}


\title{Effects of paramagnetic fluctuations on the thermochemistry of MnO (100) surfaces in the oxygen evolution reaction}

\author{Sangmoon Yoon}
\altaffiliation{Present address: Material Science and Technology Division, Oak Ridge National Laboratory, Oak Ridge, TN 37831, USA.}
\affiliation{Department of Materials Science and Engineering,
             Seoul National University, Seoul, 08826, Korea}
\affiliation{Department of Physics,
             Kyung Hee University, Seoul, 02447, Korea}
            
\author{Kyoungsuk Jin}
\altaffiliation{Present address: Department of Chemistry, Korea University, Seoul, 02841, Korea.}
\affiliation{Department of Materials Science and Engineering,
             Seoul National University, Seoul, 08826, Korea}

\author{Sangmin Lee}
\affiliation{Department of Materials Science and Engineering,
             Seoul National University, Seoul, 08826, Korea}

\author{Ki Tae Nam}
\affiliation{Department of Materials Science and Engineering,
             Seoul National University, Seoul, 08826, Korea}
             
\author{Miyoung Kim}
\email[Corresponding author. E-mail: ]{mkim@snu.ac.kr}
\affiliation{Department of Materials Science and Engineering,
             Seoul National University, Seoul, 08826, Korea}

\author{Young-Kyun Kwon}
\email[Corresponding author. E-mail: ]{ykkwon@khu.ac.kr}
\affiliation{Department of Physics,
             Kyung Hee University, Seoul, 02447, Korea}

\date{\today}

\begin{abstract}
We investigated the effects of paramagnetic (PM) fluctuations on the thermochemistry of the MnO(100) surface in the oxygen evolution reaction (OER) using the ``noncollinear magnetic sampling method \textit{plus} $U$'' (NCMSM$+U$). Various physical properties, such as the electronic structure, free energy, and charge occupation, of the MnO (100) surface in the PM state with several OER intermediates, were reckoned and compared to those in the antiferromagnetic (AFM) state. We found that PM fluctuation enhances charge transfer from a surface Mn ion to each of the intermediates and strengthens the chemical bond between them, while not altering the overall features, such as the rate determining step and resting state, in reaction pathways. The enhanced charge transfer can be attributed to the delocalized nature of valence bands observed in the PM surface. In addition, it was observed that chemical-bond enhancement depends on the intermediates, resulting in significant deviations in reaction energy barriers. Our study suggests that PM fluctuations play a significant role in the thermochemistry of chemical reactions occurring on correlated oxide surfaces.
\end{abstract}

\pacs{
71.27.+a, 
71.15.Mb,
75.20.-g,
75.30.Et
}



\maketitle

\section{Introduction}

Manganese (Mn) oxides have attracted a great deal of attention over the last few decades due to their rich redox chemistry and potential use in energy applications.~\cite{{I1-1},{I1-2}} From an industrial standpoint, applications incorporating earth-abundant and environmentally friendly Mn oxides are in high demand.~\cite{I2} The full potential of Mn oxides has gradually been realized with the development and improvement of nanosynthetic methods and surface treatment techniques.~\cite{{I3-2},{I3-1},{I3-3}} Over the last decade, nanosized Mn oxides have been widely investigated for energy conversion and storage applications, e.g. electro- and photo-catalysis for the oxygen evolution reaction (OER), supercapacitors, and lithium-ion batteries.~\cite{{I4-1},{I4-2},{I4-3}} Most of these achievements have resulted from sophisticated nano-synthesis and systematic experimental characterization.~\cite{{I5-1},{I5-2},{I5-3}} However, there is still considerable room for performance improvement by systematically and theoretically investigating chemical reactions on the Mn oxide surface.

Crystalline Mn oxides, such as MnO, Mn$_3$O$_3$, Mn$_3$O$_4$, and MnO$_2$, are paramagnetic (PM) at temperatures exceeding room temperature.~\cite{{I6-1},{I6-2}} In other words, molecules interact with surface Mn ions that have rapidly fluctuating spins. The effects of such PM fluctuations on chemical reactions occurring on the Mn oxide surface have not been fully revealed for two distinct reasons. Firstly, \textit{ab initio} calculations of PM oxides, especially in non-bulk configurations within a large supercell, pose a significant challenge. For example, density functional theory \textit{plus} $U$ (DFT+$U$) calculations require that materials be in a spin- or orbital-ordered state preventing description of spin fluctuations.~\cite{{I7-1},{I7-2}} On the other hand, DFT \textit{plus} dynamical mean field theory (DFT+DMFT) can describe spin fluctuations, but is not suitable for supercell calculations because of its dimensionality and computational costs.~\cite{{I8-1},{I8-2}} Secondly, the effects of magnetic interactions on chemical reactions have been underestimated on the order of a few meV.~\cite{{I9-2},{I9-3},{I9-1}} However, if PM fluctuations are capable of modifying electronic structure and chemical bonds, they could also have significant effects on chemical reactions. Therefore, the influence of PM fluctuations on the thermochemistry of Mn oxide surfaces is worth investigating. 

\begin{figure*}[ht]
\includegraphics[width=\textwidth]{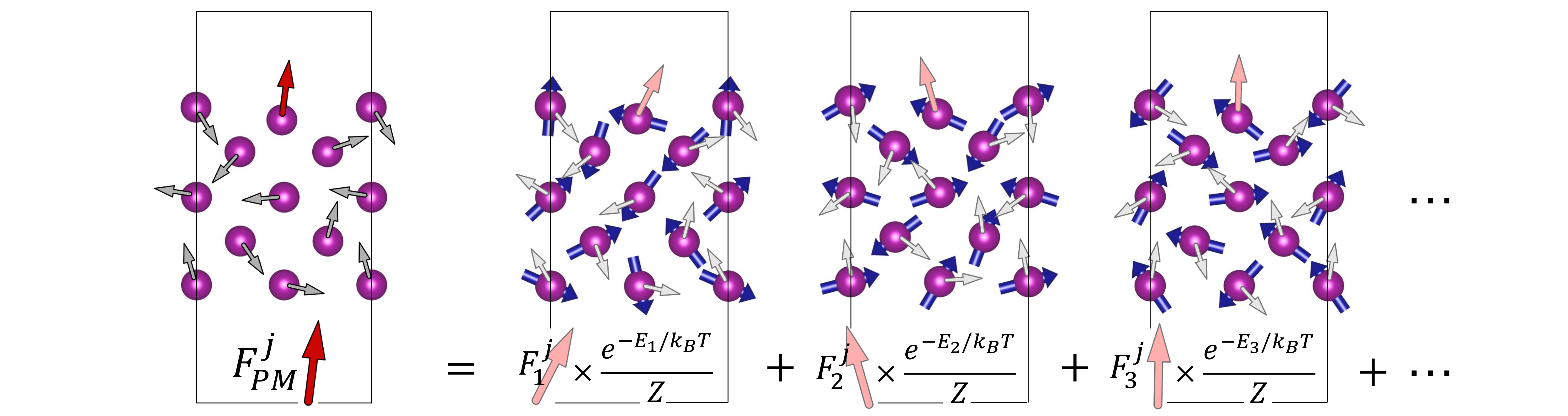}
\caption{Schematic illustration of the evaluation of the ensemble averaged Hellmann-Feynman force, used for finding the equilibrium structures of paramagnetic (PM) materials. The force exerted on each PM atom (denoted by dark red or grey arrows in the left image) is taken as the ensemble-averaged force of several individual microstates (denoted by light red or grey arrows in the three right images) using the equation shown below. Purple spheres and dark blue arrows indicate PM atoms and their local magnetic moments, respectively. See the main text for a more detailed explanation.
\label{Fig1}}
\end{figure*}

In this study, we investigated the effects of PM fluctuations on the thermochemistry of MnO (100) surfaces during the OER using the “noncollinear magnetic sampling method \textit{plus} $U$” (NCMSM+$U$)~\cite{I10}. We found that PM fluctuations facilitate charge transfer between surface Mn ions and intermediates, resulting in stronger chemical bonds with surrounding intermediates. Furthermore, the enhanced interactions caused by the PM fluctuations lead to significant deviations in the reaction energy barrier of chemical reactions.

\section{Computational methods}

Developing an \textit{ab initio} theory to describe strongly correlated materials in the PM state has been a central issue for decades. A consistent description of PM metals and Mott insulators became possible only after the development of DFT+DMFT, in which spin fluctuations are spontaneously involved through frequency-dependent self-energy.~\cite{{C1-1},{C1-2}} However, application to non-bulk configurations with large supercells, e.g. surfaces, interfaces, and structural defects, is barely feasible because of the high computational costs. Another useful approach is to perform disordered local moment (DLM) calculations combined with DFT+$U$, where spin fluctuations are considered at the static limit.~\cite{{I10},{C2-1},{C2-2}} This approach has been proven to describe PM insulators especially well, since their insulating phase is characterized by slow dynamic charge correlations such that their spin fluctuations are not affected by past spin states.~\cite{I10} In the last decade, DLM approaches have attracted considerable attention because of their applicability to the study of non-bulk configurations.~\cite{{C3-1},{C3-2}}

In this study, we used the recently proposed NCMSM+$U$ approach, which was shown to successfully explaine the PM states of Mott insulators with only moderate computational costs.~\cite{I10} In NCMSM+$U$, PM disorders can be described through a canonical ensemble of noncollinearly disordered magnetic structures, while strong electron correlations are accounted for by DFT+$U$. Magnetically disordered supercells for NCMSM+$U$ were constructed using two random number generators that independently determine the polar and azimuthal angles of each magnetic moment, with the constraint that the total magnetic moment should be zero. The electronic structure of each individual magnetically disordered microstate was explored using DFT+$U$. Then, any physical quantity in the PM phase, $X_\mathrm{PM}$, can be calculated at a given temperature $T$ by
\begin{equation}
\label{equation1}
  X_\mathrm{PM} = \frac{1}{Z}\sum_i X_ie^{{-E_i}/k_\mathrm{B}T},
\end{equation}
where $X_i$ and $E_i$ are the specific physical quantity and energy of the $i$-th microstate, respectively. Here, $Z$ defined by $\sum_ie^{-E_i/k_\mathrm{B}T}$, is a partition function. We have previously used this NCMSM$+U$ scheme to accurately evaluate the electronic profile, local magnetic moments distribution, superexchange coupling constant, and Neel temperature of PM MnO, with strong agreement with experimental results.~\cite{I10}

To explore various thermodynamic properties, it is essential to find the equilibrium structures of PM materials, which can be constructed by minimizing the forces between atoms within the multidimensional space of lattice parameters. Such forces, however, cannot be determined in PM materials due to their dynamically varying magnetic disorders. To resolve this issue, we used Eq.~(\ref{equation1}) to calculate the ensemble-averaged Hellman-Feynman forces exerted on individual atoms, which is schematically illustrated in Fig.~\ref{Fig1}, and minimize them within the multidimensional space of lattice parameters by applying a conjugate gradient algorithm. This scheme assumes that spin fluctuations and atomic motion can be separated; that is, spin fluctuation is considered to be sufficiently faster than atomic motion that atoms can be assumed to be fixed. If the time scale of spin fluctuations is similar to that of atomic motion, then ensemble-averaged forces could result in unrealistic structures. However, in most cases, spin fluctuations are regarded as one-to-two orders of magnitude faster than atomic motion at high temperature. The ensemble-averaged dynamic matrix of bcc Fe accurately reproduced its experimental high-temperature phonon bands and demonstrated the validity of ensemble-averaged forces in high-temperature atomic dynamics.~\cite{C4}

All \textit{ab initio} calculations were performed using the Vienna \textit{ab initio} simulation package (VASP)~\cite{C5}. Perdew–Burke–Ernzerhof \textit{plus} $U$ (PBE+$U$) was used to determine the exchange-correlation functional, in which double-counting interactions were corrected using the full localized limit (FLL)~\cite{I7-2}. The $U$ parameter used for Mn atoms was 4.0~eV.~\cite{I10} A plane wave basis set with a cut-off energy of 500~eV was used to expand the electronic wave functions and valence electrons were described using projector-augmented wave potentials. A symmetric MnO (100) slab supercell with 112 atoms (56 Mn and 56 O atoms) was used in this study (see Fig.~1S in Supplementary Information). The in-plane lattice constants of the supercell were fixed to the experimental value of 4.4315~\AA, while the positions of atoms were relaxed by conjugated gradient algorithms until none of the remaining ensemble-averaged Hellman-Feynman forces on any atoms exceeded 0.03~eV/\AA. A $\Gamma$-centred $4\times4\times1$ Monkhorst-Pack $k$-point grid was used for sampling the Brillouin zone. Seven noncollinear magnetically disordered microstates were used for the canonical ensemble at $T=300$~K.~\cite{I10}

\section{Results}
\subsection{Electronic structure of the PM MnO(100) surface}

\begin{figure}[t]
\includegraphics[width=1.0\columnwidth]{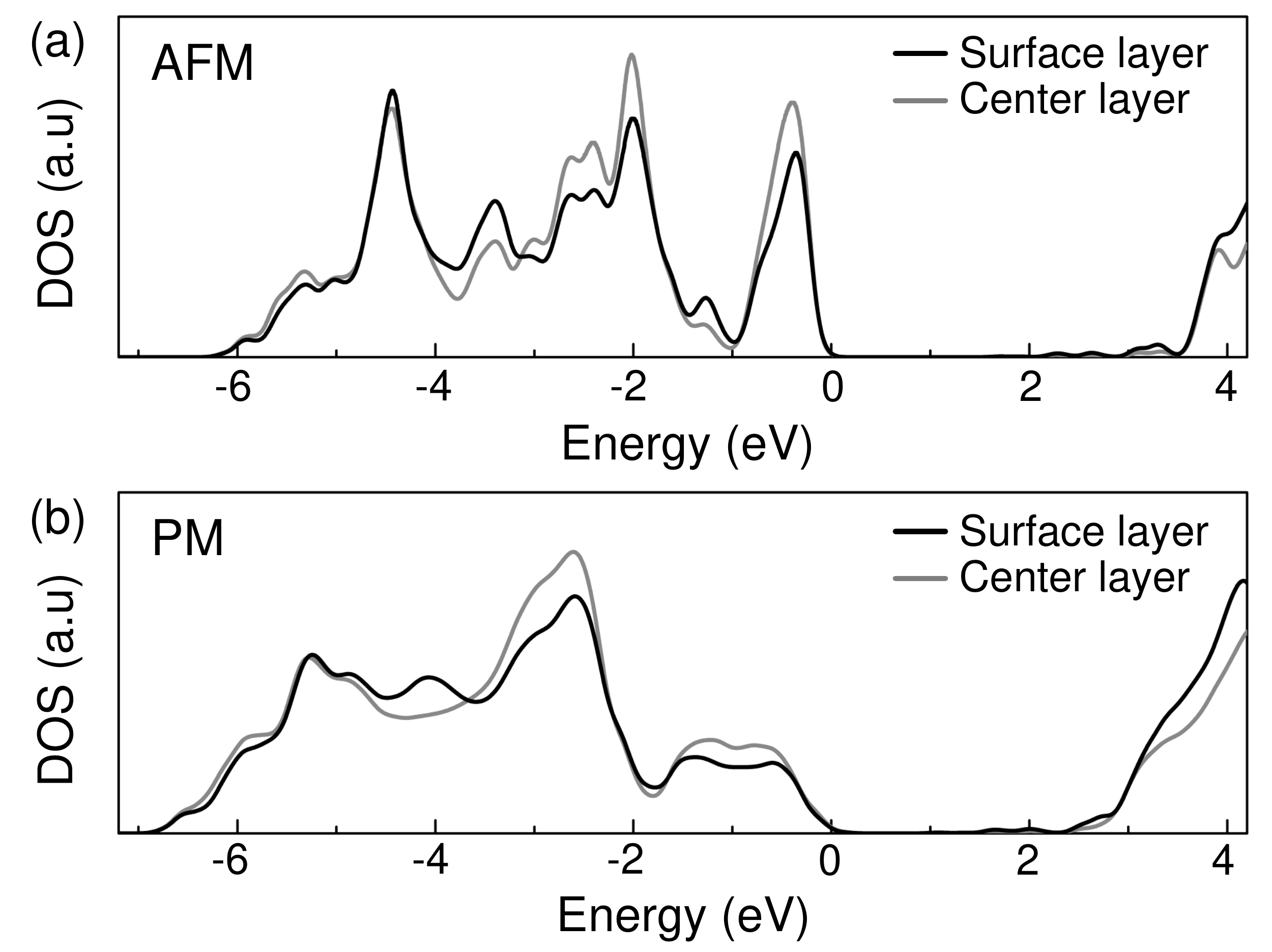}
\caption{Layer-projected local density of states (LDOS) of (a) antiferromagnetic (AFM) and (b) PM MnO slabs with a (100) surface calculated using the noncollinear magnetic sampling method \textit{plus} U (NCMSM$+U$) method with $U=4$~eV. The LDOS of the PM MnO (100) surface was estimated using the ensemble-averaged LDOS of the seven different magnetic microstates. The black and grey solid lines indicate the LDOS projected on the top and central layers, respectively.
\label{Fig2}}
\end{figure}

Figure~\ref{Fig2} shows the local density of states (LDOS) of both antiferromagnetic (AFM) and PM MnO slabs with a (100) surface projected on the surface and central layers. The LDOS for the PM phase was given by the ensemble averaged LDOS for its individual microstates. Intriguingly, NCMSM$+U$ revealed a delocalized electronic characteristic near the valence bands in the PM phase;
this has been observed in a spectral function of PM MnO calculated by DFT+DMFT~\cite{R1} and an experimental X-ray photoelectron spectroscopy (XPS) spectrum.~\cite{R1}
Note that this is an unique feature of the PM state, which cannot be simulated by conventional static DFT$+U$ calculations.
We suggest that the delocalization caused by the PM fluctuations can affect bonding with surrounding molecules and eventually modify the thermochemistry of the Mn oxide surface.
On the other hand, our calculation well described the localized feature in the counterpart AFM data.

\subsection{Effects of PM fluctuations on the thermochemistry of the MnO (100) surface}

\begin{figure*}[t]
\begin{center}
\includegraphics[width=0.95\textwidth]{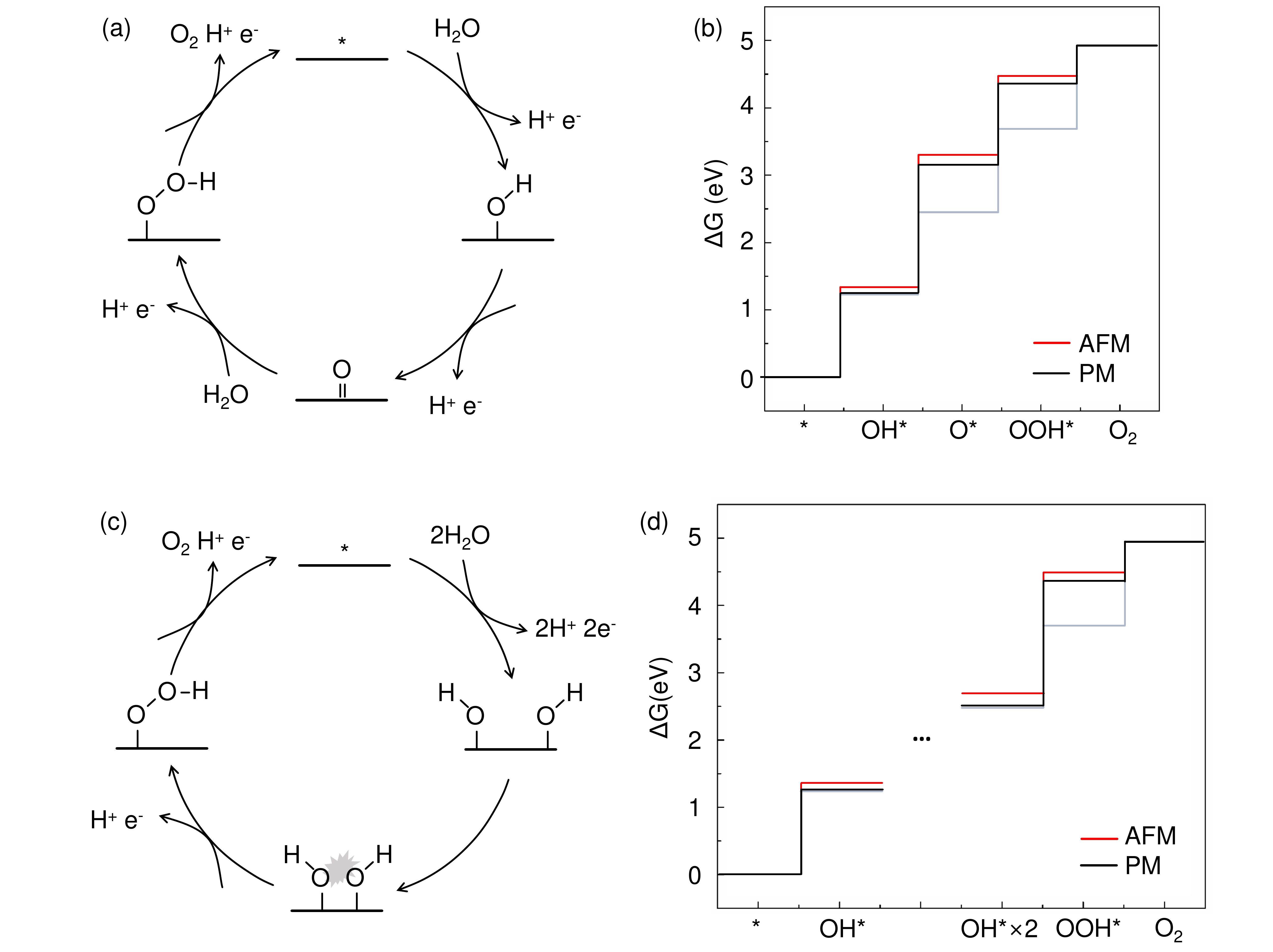}
\caption{Schematic illustration of two probable routes for the oxygen evolution reaction (OER) and corresponding free energy diagrams for the nucleophilic attack (NA) mechanism (a, b) and the direct attack (DA) mechanism (c, d). The black and red solid lines indicate the free energies of OER intermediate states occurring on the PM and AFM MnO (100) surfaces, respectively. The free energy of the intermediate state on the PM surface was given by the ensemble-averaged free energy for seven different magnetically disordered surfaces. For comparison, the free energy of an ideal OER catalyst, where each reaction step has a uniform energy barrier of 1.23~eV, corresponding to the chemical potential of the OER, is indicated by the solid grey line.
\label{Fig3}}
\end{center}
\end{figure*}

\begin{table}[b]
\caption{Free energies ($\Delta G$) of the OER intermediate states and overpotentials ($\eta$) at the AFM and PM surfaces for the nucleophilic attack (NA) and direct attack (DA) mechanisms. $\Delta G$ given by Eq.~(\ref{equation2}) is referenced to that of the pristine surface, and its values for the PM surface were evaluated by taking the ensemble-averaged value for seven different magnetically disordered configurations. $\eta$ is defined as the difference between the energy barrier of the rate-determining step (RDS) and the equilibrium potential of 1.23~eV for the OER.}
\label{table1}
\begin{tabular*}{\columnwidth}{@{\extracolsep{\fill}}ccccccc}
\hline\hline
\multicolumn{1}{c}{} & \multicolumn{5}{c}{$\Delta G$ (eV)} & $\eta$ (eV) \\
\hline
NA  & Pristine & OH* & O* & OOH* & O$_2$ & \\
\cline{2-6}
AFM  & 0 & 1.33 & 3.30 & 4.47 & 4.92 & 0.74 \\
PM   & 0 & 1.25 & 3.15 & 4.36 & 4.92 & 0.67 \\
\hline
DA  & Pristine & OH* & OH*$\times2$ & OOH* & O$_2$ \\
\cline{2-6}
AFM  & 0 & 1.33 & 2.66 & 4.47 & 4.92 & 0.58 \\
PM   & 0 & 1.25 & 2.49 & 4.36 & 4.92 & 0.64 \\
\hline\hline
\end{tabular*}
\end{table}

To investigate the effects of PM fluctuations on the thermochemistry of MnO (100) surfaces during the OER, we estimated the free energies of various OER intermediate states and compared them between the PM and AFM surfaces. Equilibrium structures for intermediate states on the PM surface were obtained using ensemble-averaged Hellman-Feynman forces, as described in detail in the Computational Methods section. The free energy of each OER intermediate step was calculated using the following equation
\begin{equation}
\label{equation2}
  \Delta G = \Delta E + \Delta{ZPE} - T\Delta S
\end{equation}
where $\Delta E$, $\Delta{ZPE}$, and $\Delta S$ are the reaction energy calculated by DFT+$U$, the difference in zero-point energy (ZPE) due to the reaction, and the change in entropy, respectively. We made use of the values of ZPEs and entropies for various OER intermediates tabulated elsewhere.~\cite{{R2-1},{R2-2}} As shown in Fig.~\ref{Fig3}, we considered two probable routes for the OER and evaluated the free energies of the intermediate states on the PM and AFM surfaces along each route; the results are summarized in Table~\ref{table1}. The two probable OER routes we considered are based on the nucleophilic attack (NA) and direct attack (DA) mechanisms~\cite{R3,S4,S5} and their reaction pathways are shown in Figs.~\ref{Fig3}(a) and (c). In the OER based on the NA mechanism, oxygen (O$_2$) molecules are generated through a process in which OH*, O*, and OOH* intermediates are formed sequentially on a monometal site. In the DA mechanism-based OER, on the other hand, O$_2$ molecules are generated through a process in which an OH* intermediate encounters another one on the surface and form an unstable OOH* species. Though Mn-based OER catalysts are widely used in the alkaline or neutral conditions,~\cite{S1,S2} the OER routes in the acid condition were examined in this study for theoretical convenience. Note that the reaction routes in the alkaline and acid conditions are equivalent from a thermodynamic perspective.~\cite{R2-1,R2-2,S3}

Note that in thermochemistry, the rate-determining step (RDS) is defined as the reaction step requiring the largest amount of energy to overcome the energy barrier, while the resting state is defined as the intermediate state immediately preceding the RDS. Figure~\ref{Fig3}(b) and Table~\ref{table1} show that, for both AFM and PM, the reaction step for evolution from OH* to O* is the RDS, and surface coverage with OH* is the resting state in the NA pathway. Our \textit{in-situ} UV experiment showed that OH* intermediates have been accumulated before O$_2$ molecules are generated on the surface of MnO micropowders (see Fig.~S1 in Supplementary Information). It further verified that the reaction step from OH* to O* is the RDS for the OER on MnO micropowders, which is consistent with our numerical results.
The overpotentials ($\eta$), determined by the difference between the energy barrier height and equilibrium potential of the reaction (1.23 eV for OER), were 0.74 and 0.67 eV at the AFM and PM surfaces, respectively. This result also implies that O* is spontaneously oxidized to OOH* once OH* evolves to O*. Thus, we included a direct reaction between OH intermediates, but not between O intermediates, as an alternative OER mechanism. As shown in Fig.~\ref{Fig3}(d) and Table~\ref{table1}, the step in which two OH* encounter each other to form an OOH* intermediate is the RDS, and the surface coverage with OH* is the resting state in the DA pathway. The value of $\eta$ for the DA pathway was 0.58 and 0.64 eV on the AFM and PM surfaces, respectively. Interestingly, $\eta$ was lower in the DA pathway than in the NA pathway. Nevertheless, the NA pathway should be regarded as the dominant reaction pathway in the OER because attacks by water molecules (H$_2$O) or hydroxyl ions (OH$^{-}$) may occur more frequently than collisions between two OH* intermediates.

\begin{table*}[ht]
\caption{Charge occupations of intermediate species and surface Mn ions. The Bader excess charge ($\Delta n_\mathrm{I}$) of the intermediate species and on-site $d$ orbital occupation ($f_\mathrm{Mn}$) of the surface Mn ions are summarized. Here, a surface Mn ion corresponds to a Mn atom that is directly connected to an intermediate at the surface. $\Delta n_\mathrm{I}$ is defined as the difference between the Bader charge of an intermediate binding to the surface and its atomic charge before binding to the surface. The left and right columns in $f_\mathrm{Mn}$ indicate the orbital occupations of the majority and minority spins, respectively, where the occupations of $d_{3z^2-r^2}$, $d_{x^2-y^2}$, $d_{yz}$, $d_{zx}$, and $d_{xy}$ orbitals are represented from top to bottom.}
\label{table2}
\begin{tabular*}{\textwidth}{cccc@{\extracolsep{\fill}}ccc}
\hline \hline
\multicolumn{3}{c}{} & Pristine & OH*   & O*    & OOH* \\ 
\hline
\multirow{2}{*}{$\Delta n_\mathrm{I}$} 
   & AFM &     &     & 0.495 & 0.681 & 0.425 \\
   & PM  &     &     & 0.548 & 0.739 & 0.554 \\ \hline
\multirow{7}{*}{$f_\mathrm{Mn}$} & AFM & 
   $\begin{matrix}
     d_{3z^2-r^2} \\ d_{x^2-y^2} \\ d_{yz} \\ d_{zx} \\ d_{xy}
   \end{matrix}$ &
   $\begin{pmatrix} 0.94 & 0.05 \\ 0.98 & 0.08 \\ 0.94 & 0.02 \\ 0.94 & 0.02 \\ 0.94 & 0.02 \end{pmatrix}$ & 
   $\begin{pmatrix} 0.51 & 0.22 \\ 0.98 & 0.11 \\ 0.95 & 0.04 \\ 0.96 & 0.06 \\ 0.94 & 0.03 \end{pmatrix}$ & 
   $\begin{pmatrix} 0.57 & 0.25 \\ 0.98 & 0.10 \\ 0.88 & 0.04 \\ 0.96 & 0.09 \\ 0.94 & 0.03 \end{pmatrix}$ & 
   $\begin{pmatrix} 0.61 & 0.18 \\ 0.98 & 0.10 \\ 0.95 & 0.05 \\ 0.95 & 0.03 \\ 0.94 & 0.03 \end{pmatrix}$ \\ \\
  & PM  & 
   $\begin{matrix}
     d_{3z^2-r^2} \\ d_{x^2-y^2} \\ d_{yz} \\ d_{zx} \\ d_{xy}
   \end{matrix}$ &
   $\begin{pmatrix} 0.94 & 0.05 \\ 0.98 & 0.08 \\ 0.94 & 0.02 \\ 0.94 & 0.02 \\ 0.94 & 0.02 \end{pmatrix}$ & 
   $\begin{pmatrix} 0.55 & 0.20 \\ 0.97 & 0.10 \\ 0.95 & 0.04 \\ 0.95 & 0.05 \\ 0.95 & 0.03 \end{pmatrix}$ & 
   $\begin{pmatrix} 0.61 & 0.24 \\ 0.97 & 0.08 \\ 0.92 & 0.09 \\ 0.93 & 0.09 \\ 0.94 & 0.02 \end{pmatrix}$ & 
  $\begin{pmatrix} 0.73 & 0.14 \\ 0.98 & 0.08 \\ 0.95 & 0.04 \\ 0.94 & 0.02 \\ 0.94 & 0.02 \end{pmatrix}$ \\ 
\hline \hline
\end{tabular*}
\end{table*}

A comparison of the free energies of intermediate states on the PM surface with those on the AFM surface reveals three important features of chemical reactions occurring on the PM surface. Firstly, the overall features of a chemical reaction do not change due to PM fluctuations; the RDS and resting state were the same on both AFM and PM surfaces. Secondly, intermediate species bind more strongly to the PM surface than to the AFM surface. The free energies of intermediate states on the PM surface are always lower than those on the AFM surface; the free energies of OH*, O*, and OOH* on the PM surface were 80, 150, and 110 meV lower, respectively, than those on the AFM surface (Table~\ref{table1}). It should be noted that energy differences of the order of 100 meV cannot be explained by magnetic interactions alone.~\cite{{I9-2},{I9-3},{I9-1}} The PM fluctuations that modify chemical bonds may contribute significantly to such differences. Thirdly, the variation in interaction enhancement among the intermediate species leads to quantitative deviation of reaction energy barriers. In the NA pathway, the $\eta$ value of the PM surface is 70~meV lower than that of the AFM surface, while for the DA route, the PM surface has a $\eta$ value 60~meV higher than that of the AFM surface. Our results showed that $\eta$ can deviate by tens of meV between the PM and AFM surface, although it should be noted that deviations in $\eta$ due to PM fluctuations do not indicate a universal decrease or increase for different OER pathways. However, such deviations could be significant in the OER because differences of the order of 100 meV can determine the usefulness of a catalyst.~\cite{R4}

\subsection{Influence of PM fluctuations on charge transfer and chemical bonds}

We further analysed the charge occupations of intermediates and surface Mn ions to understand the influence of PM fluctuations on chemical bonding with intermediates (Table~\ref{table2}). It was found that each intermediate receives electrons from the surface Mn atoms as indicated by its Bader excess charge ($\Delta n_\mathrm{I}$) of the intermediate on either the AFM or PM MnO(100) surface.
Here, $\Delta n_\mathrm{I}$ is defined as the difference between the Bader charge of an intermediate binding to the surface and the valence charge of its unbound counterpart before binding to the surface.
The most important observation is the fact that all the intermediates receives more electrons from the Mn atoms on the PM surface than on the AFM surface. As shown in Table~\ref{table2}, all of the $\Delta n_{I}$ values on the PM surface are larger than the corresponding values found for the AFM surface. This indicates that charge transfer from the surface Mn ion to intermediates is significantly enhanced on the PM surface. 
Moreover, the $\Delta n_{I}$ values for OH* and OOH* intermediates are smaller than that for the O* intermediate, reflecting the trend in the formal charges of the intermediates which are $1^-$ for OH* and OOH*, and $2^-$ for O*. Our analysis of the orbital occupations of surface Mn ions ($f_\mathrm{Mn}$) revealed that the $d_{3z^2-r^2}$ orbital mainly contributes to electron transfer when binding with OH* and OOH*, while for chemical bonding with O*, there is a small contribution from the $d_{yz}$ or $d_{zx}$ orbital in addition to the major contribution from the $d_{3z^2-r^2}$ orbital.

\section{Discussion}

Our LDOS, free energy, and site-dependent charge analyses yielded an important insight into the role of PM fluctuations in the thermochemistry of the MnO (100) surface. Bader charge analyses showed that more electrons are transferred from Mn ions to intermediates on PM surfaces. From the LDOSs, we observed that valence bands are delocalized due to PM fluctuations. This delocalization is believed to be a cause of the enhanced charge transfer. In ionic bonding materials, more charge transfer results in stronger bonds, that is, an enhancement in charge transfer at the PM surface could affect the strength of the bonds between surface atoms and reaction intermediates. This study verified the influence of PM fluctuations on the thermochemistry of a MnO(100) surface during the OER. Delocalization of electronic states due to PM fluctuations is one of the universal features of correlated PM phases oxides.~\cite{{D1-1},{D1-2},{D1-3}} We therefore expect that similar effects of PM fluctuations on OER thermochemistry will exist in other chemical reactions on correlated oxide surfaces. 

\section{Conclusions}

We employed the NCMSM+$U$ approach~\cite{I10} to investigate the effects of PM fluctuations on the thermochemistry of MnO (100) surfaces during the OER. The equilibrium structures, free energies, electron occupations, and charge transfers of the PM MnO (100) surface with intermediates in the PM state were calculated using NCMSM+$U$ and compared to those on the AFM surface. It was found that PM fluctuation strengthens the chemical bonding of intermediates to the MnO surface through enhancement of the charge transfer between them, which can be attributed to the delocalized nature of the valence bands. Nonetheless, this effect does not alter the overall features, such as the RDS and resting state, of two possible reaction pathways for the OER. We observed that the dependence of such interaction enhancement on different kinds of intermediates results in a significant deviation of the reaction energy barrier of the order of 100~meV, which may be of critical importance in various chemical reactions including the OER. Our findings suggest that when investigating chemical reactions quantitatively, PM fluctuation should be taken into account, because it directly influences chemical interactions of participating molecules to PM surfaces. Furthermore, our study demonstrated the applicability of DLM approaches, including the NCMSM$+U$ approach, to \textit{ab initio} studies of the thermochemistry of PM energy materials.

\section*{Conflicts of interest}
There are no conflicts to declare.

\section*{Acknowledgements}

We acknowledge financial support from the Korean government through National Research Foundation (2017R1A2B3011629, 2019R1A2C1005417), and Korea Evaluation Institute of Industrial Technology (K\_G011008062504). Some portion of our computational work was done using the resources of the KISTI Supercomputing Center (KSC-2018-CHA005 and KSC-2020-CRE-0011).

\bibliographystyle{apsrev}
\bibliography{Reference.bib} 
\end{document}




\title{Supporting Information for \\
Effects of paramagnetic fluctuations on the thermochemistry of MnO (100) surfaces in the oxygen evolution reaction} 



\author{Sangmoon Yoon}
\affiliation{Department of Materials Science and Engineering,
             Seoul National University, Seoul, 08826, Korea}
\affiliation{Department of Physics,
             Kyung Hee University, Seoul, 02447, Korea}

\author{Kyoungsuk Jin}
\affiliation{Department of Materials Science and Engineering,
             Seoul National University, Seoul, 08826, Korea}

\author{Sangmin Lee}
\affiliation{Department of Materials Science and Engineering,
             Seoul National University, Seoul, 08826, Korea}

\author{Ki Tae Nam}
\affiliation{Department of Materials Science and Engineering,
             Seoul National University, Seoul, 08826, Korea}
             
\author{Miyoung Kim}
\email[Corresponding author. E-mail: ]{mkim@snu.ac.kr}
\affiliation{Department of Materials Science and Engineering,
             Seoul National University, Seoul, 08826, Korea}

\author{Young-Kyun Kwon}
\email[Corresponding author. E-mail: ]{ykkwon@khu.ac.kr}
\affiliation{Department of Physics,
             Kyung Hee University, Seoul, 02447, Korea}
             
\maketitle

\begin{figure}[h]
\includegraphics[width=1.0\columnwidth]{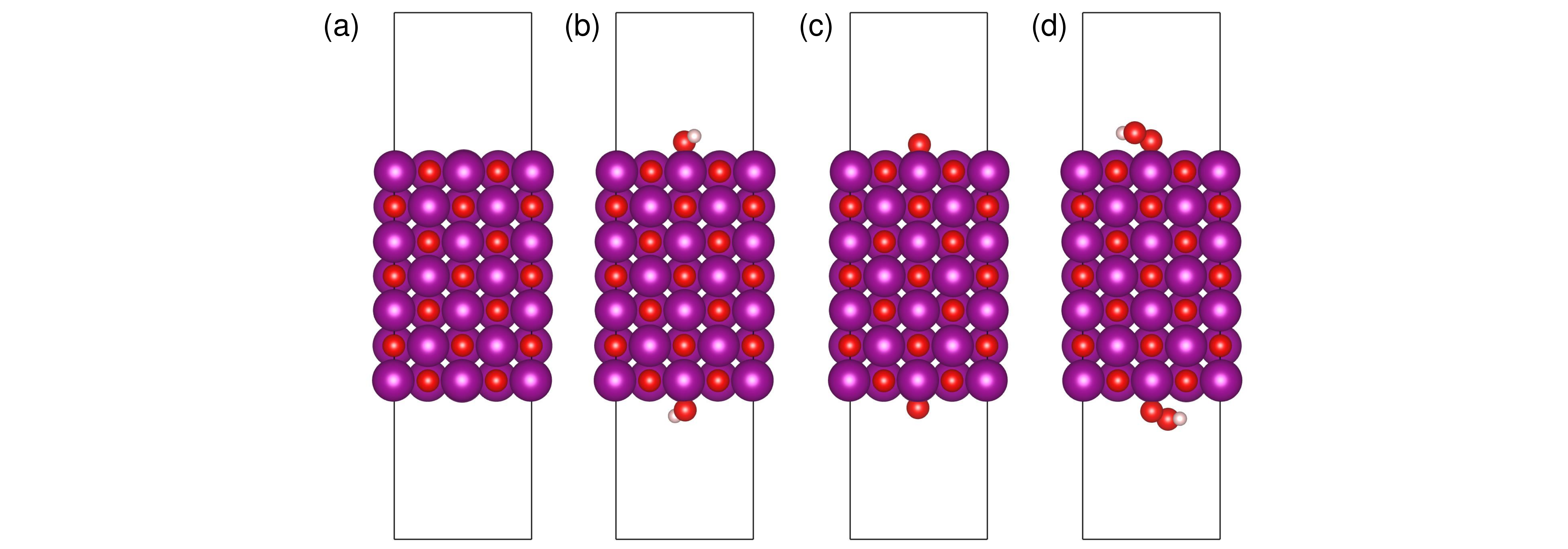}
\caption{(Color online) Symmetric slab models used to calculate the Gibbs free energies of intermediate states in the OER: (a) pristine surface model, and (b)-(d) surface models with OH*, O*, and OOH* intermediates, respectively. The thickness of vacuum layers between slabs is about 15~{\AA}.}
\end{figure}


\begin{figure}[h]
\includegraphics[width=1.0\columnwidth]{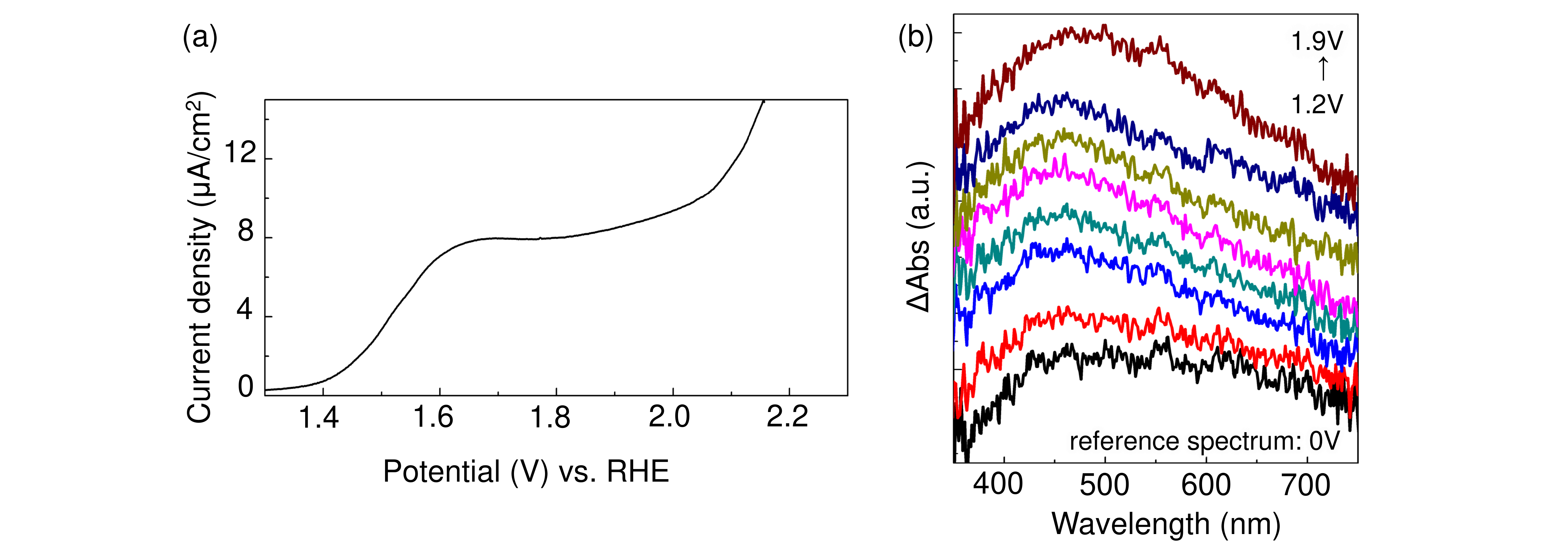}
\caption{(Color online) Experimental evidence of the OH* resting state for the OER in MnO micropowder. (a) Polarization-corrected CV curve of MnO powders in 0.5~M phosphate buffer at pH~7. (b) \textit{In-situ} UV difference spectra of MnO powder with increasing potential from 1.2~V to 1.9~V. A spectrum measured at the open circuit voltage was used as the reference spectrum.}
\end{figure}

To confirm the validity of our calculations, we performed two experiments with commercial MnO micropowder, which is expected to have stable surfaces. Figure~S2(a) shows the polarization-corrected cyclic voltammetry (CV) curve of the MnO micropowder at pH~7. A clear additional redox reaction peak can be seen at approximately 1.6~V (vs.~RHE) before the Faraday current appears. Figure~S2(b) shows the \textit{in-situ} UV-visible differential spectra of the MnO micropowder. Each differential spectrum was obtained by subtracting the reference spectrum from that obtained at each potential. As the potential increased from 1.2~V to 1.9~V, a broad peak at around $450-500$~nm appears from 1.4~V. It has been reported previously that this feature indicates the formation of Mn$^{3+}-$~OH species on the Mn oxide surface.$^{1,2}$ This shows that, prior to 1.9~V, i.e. at the onset potential for oxygen evolution, Mn$^{3+}-$~OH species accumulated on the surface of MnO micropowder. Both experimental results verify that a redox reaction that transforms Mn$^{2+}$+H$_2$O to Mn$^{3+}-$~OH* occurred prior to the RDS; the surface state of Mn$^{3+}-$~OH* was maintained from 1.4~V to 1.9~V, which is consistent with our DFT results. On the other hand, the spectral signature of Mn$^{4+}=$O species, a sharp peak at 400~nm with a broad peak at 700~nm, has not been observed in the UV difference spectra of MnO micropowders.$^{3}$ This result shows that the reaction step evolving from an OH* intermediate to an O* intermediate is the RDS for the OER on MnO micropowders. MnO micropowders are believed to be composed of low-index facets with much less structural defects than MnO nanoparticles.

\section*{References}
 \noindent 1 \hspace{6pt}  T. Takashima \textit{et al}. \textit{Journal of the American Chemical Society}, 2012, \textbf{134}, 1519-1527.\\
 \noindent 2 \hspace{6pt}  T. Takashima \textit{et al}. \textit{Journal of the American Chemical Society}, 2012, \textbf{134}, 18153-18156.\\
 \noindent 3 \hspace{6pt}  K. Jin \textit{et al}. \textit{Journal of the American Chemical Society}, 2017, \textbf{139}, 2277-2285.